\documentclass[useAMS,usenatbib]{mn2e}
\usepackage{times}
\usepackage{epsfig}
\usepackage{colordvi}
\usepackage{graphicx}
\usepackage{amssymb}

\def\araa{ARA\&A}
\def\apj{ApJ}
\def\apjl{ApJ}
\def\apjs{ApJS}
\def\aap{A\&A}
\def\mnras{MNRAS}

\def\aaps{A\&AS}

\title[Hybrid  Comptonization model for accreting SMBHs]{A self-consistent hybrid 
Comptonization model for broad-band spectra of accreting supermassive black holes}

\author[Alexandra Veledina, Indrek Vurm \& Juri Poutanen]
    {Alexandra Veledina,$^{1}$    Indrek Vurm$^{1,2,3}$    and     Juri~Poutanen$^{1}$ \\
$^1$Astronomy Division, Department of Physics, PO Box 3000, FIN-90014 University of Oulu, Finland \\
$^2$Racah Institute of Physics, Hebrew University of Jerusalem, 91904 Jerusalem, Israel \\
$^3$Tartu Observatory, 61602 T\~{o}ravere, Tartumaa, Estonia}

\begin{document}
\date{}
\pagerange{\pageref{firstpage}--\pageref{lastpage}} \pubyear{2010}

\maketitle

\label{firstpage}

\begin{abstract}
The nature of the broad-band spectra of supermassive accreting black holes in active galactic nuclei (AGNs) is still unknown. 
The hard X-ray spectra of Seyferts as well as of  Galactic stellar-mass black holes (GBHs) are well represented by thermal Comptonization, 
but the origin of the seed photons is less certain. 
The MeV tails observed in GBHs provide evidence in favour of non-thermal electron tails 
and it is possible that such electrons are also present in the X-ray emitting regions of AGNs.

Using simulations with the kinetic code that self-consistently models electron and photon distributions, we find that the
power-law-like X-ray spectra in AGNs can be explained in terms of the synchrotron self-Compton radiation of 
hybrid thermal/non-thermal electrons, similarly to the hard/low state of GBHs. 
Under a very broad range of parameters the model predicts a rather narrow distribution of 
photon spectral slopes consistent with that observed from LINERs and 
Seyferts at luminosities less than 3 per cent of the Eddington luminosity. 
The entire infrared to X-ray spectrum of these objects can be described in terms of our model, suggesting a tight correlation
between the two energy bands.
We show that the recently found correlation between slope and the Eddington ratio at higher  luminosities 
can be described by the increasing fraction of disc photons in the emitting region, 
which may be associated with the decreasing inner radius of the optically thick accretion disc.
The increasing flux of soft photons is also responsible for the transformation of the electron distribution from nearly thermal 
to almost completely non-thermal.
The softer X-ray spectra observed in narrow-line Seyfert galaxies may correspond to non-thermal Comptonization of the disc
photons, predicting that no cutoff should be observed up to MeV energies in these sources, similarly to the soft-state GBHs. 
\end{abstract}


\section{Introduction}

Accreting radio-quiet supermassive black holes (SMBHs) residing in the centres of quasars, Seyfert galaxies, narrow-line
Seyfert 1 galaxies (NLSy1) and some fraction of low-ionization nuclear emission-line regions (LINERs) in many respects are
analogous to the Galactic stellar-mass black holes (GBHs) in X-ray binaries. 
In the X-ray/soft $\gamma$-ray band, the spectra of Seyferts can be represented by a sum of a power-law-like continuum which
cuts off at a few hundred keV and a reflection component with the  iron fluorescent $\rm K{\rm\alpha}$ line believed to be
produced by reprocessing of the intrinsic power-law by cold opaque matter, probably the accretion disc \citep{NanPoun94}.
The 2--10 keV intrinsic spectral energy slope (defined as $F_E \propto E^{-\alpha}$) of the power-law $\alpha \sim$ 0.9--1.0,
which is ubiquitously found in Seyferts, is somewhat larger than what is measured in GBHs in their hard state $\alpha \sim$
0.6--0.8 \citep*[e.g.][]{ZLS99}. 
NLSy1 having softer X-ray spectra than Seyferts, probably represent a state with higher accretion rate (in Eddington units) 
similar to the soft/very high state of GBHs \citep*{PDO95}. 

The seemingly similar X-ray slopes of Seyferts triggered the efforts to find a physical model which would explain such spectral 
stability. The non-thermal models became popular \citep[see][ for a review]{Sve94}. These invoke injection of high-energy leptons
into the system with subsequent Compton cooling by accretion disc photons and photon-photon pair productions initiating pair
cascades. Saturated cascades produce intrinsic spectra with $\alpha \approx0.9$ \citep{ZL85,Sve87}, which becomes consistent
with the X-ray spectra of Seyferts after accounting for the hardening due to Compton reflection \citep{GF91}. The
pair-cascade models, on the other hand, predict a strong tail above 300 keV and an annihilation line which have never been
observed.   
The \textit{CGRO}/OSSE observations of the brightest Seyfert 1 galaxy NGC 4151 constrain the fraction of energy going to
non-thermal injection to be less than 50\%, while the rest of the power going to thermal heating \citep*{ZLM93,ZJM96}.  
The existing upper limits on the average flux of Seyferts above 100 keV are compatible with the presence of weak non-thermal
tails \citep[e.g.][]{Gon96, Johns97}. The more recent \textit{INTEGRAL} observations \citep{LZ10} have not improved those
constraints. 

Non-detectable high-energy tail in Seyfert galaxies can also be interpreted in terms of pure thermal models, where 
the power is equally shared among all the thermal particles. The spectra indeed can be well described by  Comptonization on
\textit{thermal} 50--150 keV electrons \citep[see e.g.][]{ZJP97,P98,Zdz99}. 
The stability of the spectral slopes can be interpreted as the evidence
of the radiative feedback between the hot X-ray emitting plasma and the cool accretion disc, which is assumed to be a sole source
of the seed photons for Comptonization. 
The slab-corona model, where the hot plasma sandwiches the accretion disc  \citep[see e.g.][]{HM91, HM93}, predicts too soft
spectra due to large flux of the reprocessed UV photons \citep{SPS95}. 
\citet*{MDM05} showed that the hard spectrum can be achieved if one assumes large ionization parameter of the disc, but in this
case the model fails on the predicted reflection properties.
Localized  active regions atop of a cool disc \citep*{HMG94,SPS95,PS96,Sve96}
are more photon starved and can produce
much harder spectra in agreement with observations. However, in this case it is difficult to understand why there is a
preferential slope, as it is a strong function of the separation of the active region from the disc. 

The detection of the MeV tails in the spectra of the GBH Cyg X-1 in both hard and soft states \citep{McConnell02} imply the
presence of a significant non-thermal component in the electron distribution.  
The power-law looking spectra extending to the MeV energies in other soft-state GBHs \citep{Grove98,ZGP01} are consistent with
being produced by non-thermal Comptonization \citep{PC98,P98}.
By analogy, the electrons in Seyferts also might have a non-thermal population. 
More probably in both types of sources, the electron distribution is hybrid in both hard and soft states \citep{PC98,P98,coppi99},
with the non-thermal fraction increasing for softer spectra.
However, non-thermal particles could be spatially separated from the thermal ones.

The presence of the non-thermal particles, even if they are not energetically dominant, has a strong impact on the emitted
spectrum, because the synchrotron emission can increase by orders of magnitude even if only 1 per cent of electron energy is in
the power-law tail \citep{WZ01}. 
As most of the synchrotron emission is self-absorbed, this process plays an important role in shaping the electron
distribution by thermalizing particles via emission and absorption of synchrotron photons, the so called synchrotron boiler 
\citep*{GGS88,GHS98}.  

It was recently shown that under the conditions of GBHs, Coulomb collisions and the synchrotron boiler efficiently thermalize
electrons producing thermal population at low energies even if the electrons were originally non-thermal \citep{VP08,PV09, MB09}.
The hard spectrum of GBHs can be fully accounted for by the synchrotron self-Compton (SSC) mechanism in the resulting hybrid
electrons. In this picture the disc photons are not required  to interact with the hot X-ray emitting plasma at all. 
This mechanism also gives a rather stable X-ray slopes for a large range of parameters, relieving the need for the feedback
between the disc and the active region.  
 
As many of the radiative processes depend only on the compactness of the source, but not on its luminosity or size separately, one
may think that the spectra of Seyferts can also be described in terms of hybrid Comptonization models. However, not all processes
can be scaled away. 
The aim of this paper is to study in details the spectral formation in relativistic hybrid plasmas in the vicinity of a
SMBH. One of our goals is to learn how the results scale with the mass of the compact object. We also test the role of
bremsstrahlung in SSC models and, finally, we compare the model with the data on Seyferts and NLSy1s.

\section{Model description}

\subsection{Geometry}

It is commonly accepted that the main process responsible for X-ray spectra of radio-quiet active galactic nuclei (AGNs) is
(nearly) the thermal Comptonization of soft seed photons by energetic electrons. The nature of the seed photons is less
understood.
The two main candidates are the standard cool accretion disc \citep{SS73}  and the synchrotron emission from the hot electron
populations. We model both types of seed photon sources.

The geometry of the X-ray emitting region also remains unclear. 
It can be the inner hot part of the accretion flow which can be radiatively inefficient as well as efficient \citep*[see reviews
in][]{NMQ98,Yuan07rev}. 
The disc-corona models are also popular (\citealt*{GRV79}; \citealt{HM91,HM93,HMG94, SPS95,PS96,P98}). 
The slab (plane-parallel) corona \citep{HM91,HM93} is unlikely to be realized, as it produces Comptonization spectra 
which are too soft compared to the observed $\alpha \sim 1$ \citep{SPS95}. 
The corona can also be patchy and consist of isolated active regions atop a cool accretion disc \citep{GRV79,HMG94,SPS95}. This
kind of geometry can produce spectra of almost arbitrary hardness if only soft photon reprocessed in the disc are considered. 
The corona does not need to be static, but can have substantial bulk velocity away from the disc or towards it \citep{B99PE}. 
In this work we consider spherically symmetric region, which can be either the inner hot part of accretion flow or a blob
of gas above the accretion disc.
No anisotropy in geometry or seed photons is taken into account.

\subsection{Main parameters}

In most of the simulations we assumed that the active region has a size of $R = 10R_{\rm S}$, where 
$R_{\rm S} = 2G M_{\rm BH}/c^2$ is the Schwarzschild radius of the black hole of mass $M_{\rm BH}$, as one can expect most of 
the energy to be liberated within a region of a corresponding size.
It is convenient to introduce a compactness parameter, which is independent of the black hole mass:
\begin{equation}
l_{\rm inj} = \frac{L_{\rm inj}}{R} \frac{\sigma_{\rm T}}{m_{\rm e} c^3} \simeq 10^3 \frac{L_{\rm inj}}{L_{\rm Edd}}  \frac{10
R_{\rm S}}{R}, 
\end{equation}
where $L_{\rm Edd} = 4\pi \mu_{\rm e} G M_{\rm BH} m_{\rm p} c / \sigma_{\rm T} = 1.3 \times 10^{38} M_{\rm BH}/\rm{M_{\odot}}$ erg
s$^{-1}$, $\sigma_{\rm T}$ is the Thomson cross-section, $\mu_{\rm e} =2/(1+X)$, and $X$ is the hydrogen fraction. 
 
The importance of synchrotron processes is determined by magnetization parameter $\eta_{\rm B} = U_{\rm B} R^2 c/ L_{\rm inj}$,
where $U_{\rm B} = B^2/(8\pi)$ is the magnetic energy density and 
$\displaystyle L_{\rm inj} \approx \frac{4\pi}{3} R^2 c U_{\rm rad}$ ($U_{\rm rad}$ is
the radiation energy density). The equipartition of magnetic and radiation energy densities occurs at $\eta_{\rm B} \approx 0.25$.
The magnetic field can be expressed in terms of adopted parameters as
\begin{equation}\label{eq:magfield}
 B = \left( \frac {8 \pi m_{\rm e} c^2}{\sigma_{\rm T}} \frac {\eta_{\rm B}\ l_{\rm inj} }{R} \right)^{1/2} \approx 
 10^3 \left( \frac {\eta_{\rm B}\ l_{\rm inj} }{M_7 \ R/10R_{\rm S} } \right)^{1/2} \mbox{G} ,
\end{equation}
where $M_7=M_{\rm BH}/10^7 \rm{M_{\odot}}$.

We assume that the released energy is given to the electrons.
The transfer of energy is modelled as a power-law injection: $dN_{\rm e}/(dt\ d\gamma) \propto \gamma^{-\Gamma_{\rm inj}}$
extending from low-energy Lorentz factor $\gamma_{\rm le} = 1.0$ to high-energy cut-off at $\gamma_{\rm he} = 10^3$.
The Thomson optical depth of the electrons is chosen at the beginning and fixed during each simulation, i.e. the number of
particles is conserved.
The injected electrons increase the optical depth; therefore, to keep it constant throughout the simulations, the same amount of
electrons is removed from the system.
This action simulates the effect of re-acceleration.
The  total luminosity of the system $L$ is equal to the net power 
$L_{\rm inj} = \frac{4\pi}{3} R^3 \dot{N}_{\rm e} (\overline{\gamma}_{\rm inj} - \overline{\gamma}_{\rm eq}) m_{\rm e} c^2$, 
where $\overline{\gamma}_{\rm inj}$ and $\overline{\gamma}_{\rm eq}$ are the mean Lorentz factors of the injected and the
equilibrium distributions. 
 
We also consider an additional source of soft photons from the accretion disc in the form of the blackbody radiation 
injected homogeneously to the system. The parameters that describe them are the temperature $T_{\rm d}$ 
and the ratio of the disc luminosity to the injected power  $f=L_{\rm disc}/L_{\rm inj}$.

Thus the main parameters of the model are the mass of the compact object, $M_{\rm BH}$, size of the active region, 
$R/R_{\rm S}$, magnetization parameter, $\eta_{\rm B}$, Eddington ratio, $L/L_{\rm Edd}$, ratio $f$ of the disc seed
photon luminosity to the injected luminosity, injection slope $\Gamma_{\rm inj}$ and the Thomson optical depth
$\tau=R \sigma_{\rm T} n_{\rm e}$.

\subsection{Characteristic time-scales}
 
There are three time-scales important in our simulations: light-crossing time $t_{\rm lc}$, cooling time $t_{\rm cool}$ and
re-acceleration time $t_{\rm ra}$.

The re-acceleration time is calculated for a given optical depth and required injection energy as
\begin{equation}
 t_{\rm ra} = \frac {4 \pi}{3 } \frac {R}{c} \frac {\tau \overline{\gamma}_{\rm inj}}{l_{\rm inj}},
\end{equation}
corresponding to the average time of the removal of the electrons from the system (as modelled in the code).
For large optical depth, the re-acceleration time is larger than the light-crossing time and thus it can be associated with the
escape time.
In the case of low optical depth, $t_{\rm ra} < t_{\rm lc}$, and it cannot be thought of as the escape time itself, but
interpreted as the time between re-accelerations.

The synchrotron cooling time can be calculated as 
\begin{equation}
 t_{\rm cool, s} \equiv \frac {\gamma - 1}{|\dot{\gamma}_{\rm s}|} = 
 \frac {1}{\gamma + 1} \left( \frac{4}{3} \frac {\sigma_{\rm T} U_{\rm B}}{m_{\rm e}c} \right)^{-1} = 
 \frac {1}{\gamma + 1}  \frac {3}{4}  \frac{R/c}{\eta_{\rm B} l_{\rm inj}} ,
\end{equation}
where $\dot{\gamma_{\rm s}}$ is the synchrotron cooling rate. 
Similarly, one can determine the cooling time by Compton scattering \citep[e.g.][]{RL79}
\begin{equation}
 t_{\rm cool, cs} \equiv \frac {\gamma - 1}{|\dot{\gamma}_{\rm cs}|} = 
 \frac {1}{\gamma + 1} \left( \frac{4}{3} \frac {\sigma_{\rm T} U_{\rm rad}}{m_{\rm e}c} \right)^{-1} = 
 \frac {1}{\gamma + 1}  \frac{\pi R/c}{l_{\rm inj}}.
\end{equation}
The ratio of the two time-scales is roughly
\begin{equation}\label{eq:cools/cs}
\frac{t_{\rm cool, s}}{t_{\rm cool, cs} } = \frac {U_{\rm rad}}{U_{\rm B}} =  \frac{3}{4 \pi \eta_{\rm B} }.
\end{equation}
For most of the simulations, the synchrotron cooling time is comparable to (or more than) the Compton cooling time.

The e-e Coulomb losses at the equilibrium can be estimated as \citep[e.g.][]{NM98} 
\begin{equation}
 t_{\rm cool, Coul} \equiv \frac {\gamma - 1}{|\dot{\gamma}_{\rm Coul}|} \approx 
                    \frac {2 R}{3 c} \frac{\overline{\gamma}_{\rm eq}} {\tau \ln \Lambda} \frac {p^3}{\gamma(\gamma+1)},
\end{equation}
where $\ln \Lambda$ is the Coulomb logarithm (we take $\ln \Lambda = 16$).
The bremsstrahlung cooling time is larger (approximately by the inverse of the fine structure constant 
$1/\alpha_{\rm fs}\approx137$).
At higher energies, the cooling is determined by radiative processes, while for lower energies non-radiative Coulomb collisions
are dominant. The relevant scaling is
\begin{equation}\label{eq:Coul/cs}
 \frac {t_{\rm cool, Coul}}{t_{\rm cool, cs}} \approx \frac{2 \overline{\gamma}_{\rm eq}}{3 \pi \ln \Lambda}
                                             \frac{p^3}{\gamma} \frac{l_{\rm inj}}{\tau}
\end{equation}

The ratio of the synchrotron cooling and re-acceleration times can be approximated as ($\overline{\gamma}_{\rm inj} \approx 2$ in
our simulations):
\begin{equation}\label{eq:cool/ra}
\frac{ t_{\rm cool, s}}{t_{\rm ra} } \approx \frac {0.1}{\tau \eta_{\rm B}} \frac {1}{\gamma+1},
\end{equation}
and the ratio of the Coulomb cooling to re-acceleration time can be written as 
\begin{equation}\label{eq:Coul/ra}
\frac{ t_{\rm cool, Coul}}{t_{\rm ra} } \approx 5.6 \frac {\overline{\gamma}_{\rm eq} p^3}{\tau^2 \gamma(\gamma+1)}
 					\left( \frac {L}{L_{\rm Edd}} \right) \left( \frac {R}{10R_{\rm S}}\right)^{-1}.
\end{equation}
As long as the optical depth is not too low, the average cooling time is less than the re-acceleration time. This condition is
satisfied for the majority of simulations presented in this paper.
 
 \subsection{Radiative processes and numerical simulations}

The gravitational energy released as matter goes down the potential well of the black hole heats the protons and can be
transferred to electrons via various mechanisms, for instance, Coulomb collisions with protons, magnetic reconnection,
collective plasma effects or shock acceleration.
The accelerated electrons subsequently cool, producing a radiation field in the region.
The presence of the latter strongly affects the electron distribution. Due to acceleration processes operating in the system,
the electron distribution might not be Maxwellian.
Even a small non-thermal tail (with the energy content of only 1 per cent) can significantly increase the net (after accounting
for self-absorption) synchrotron emission up to $\sim 10^5$ times \citep{WZ01} and therefore affect the spectrum of the
Comptonized component. 

For a given particle distribution (Maxwellian, power-law or hybrid) one can easily calculate the emission spectrum, but the former
in turn depends on the radiation field. Thus, to find an equilibrium spectrum, which is self-consistent with the electron
distribution, one has to solve coupled time-dependent kinetic equations for all the particle species present in the system. 

Formal equilibrium solution of the kinetic equation for electrons, when cooling dominates over the escape, can be written as
\citep{BG70}
\begin{equation}\label{eq:formsol}
 \frac {d N_{\rm eq}}{d\gamma} \propto \frac {1}{\dot{\gamma}} 
                    \int \limits_{\gamma}^{\gamma_{\rm he}} \gamma'^{-\Gamma_{\rm inj}} d \gamma',
\end{equation}
where $d N_{\rm eq}/d\gamma$ represents the equilibrium particle distribution and $\dot{\gamma}$ is the sum of all possible
cooling rates. 
In the radiative (Compton and synchrotron) cooling dominated regime, $\dot{\gamma} \propto \gamma^2$, while if the electrons
exchange energies predominantly by Coulomb collisions, then $\dot{\gamma} \propto \gamma^0$.
Substituting the cooling rates into equation~(\ref{eq:formsol}) and integrating 
(hereinafter we consider $\Gamma_{\rm inj} \ge 2$), one gets the equilibrium electron distribution in the form 
$d N_{\rm eq}/d\gamma \propto \gamma^{-s}$ with
 \begin{equation}\label{eq:elslope}
 s = \left\{
 \begin{array}{ccc}
  \Gamma_{\rm inj} + 1, & \dot{\gamma} \propto \gamma^2 & {\rm (Compton,\,synchrotron)} \\
  \Gamma_{\rm inj} - 1, & \dot{\gamma} \propto \gamma^0 & {\rm (Coulomb)}.
 \end{array} \right.
 \end{equation}
In a general case, it is necessary to account for particle heating and thermalization; therefore, it is impossible to find an
analytical solution; the problem has to be solved numerically. 

We consider electrons, positrons and photons. Neutrality of the plasma is provided by protons, which are assumed to be at rest and
participate only in electron-proton bremsstrahlung. The following elementary processes are taken into account. The particles
interact with each other via Coulomb (M{\o}ller/Bha-bha) collisions.
The cooling/heating of the electrons (and positrons) occurs by emitting and absorbing synchrotron radiation, bremsstrahlung
emission (electron-proton, electron-electron, positron-positron, electron-positron) and Compton scattering.
The seed soft photons for Compton scattering can be provided by the synchrotron as well as by the accretion disc. 
Photon-photon pair production and pair annihilation are also taken into account; however, these processes play a minor role for
considered parameters.

We assume particle and photon distributions to be homogeneous and isotropic. 
The calculations are done in a one-zone geometry with tangled magnetic field.
The photon escape from the region is modelled by the escape probability formalism.
The escaping photon distribution can be described by the differential luminosity $L_E=dL/dE$ or 
differential compactness $l_{E}=(l_{\rm inj}/L) \ L_E$. 

To obtain the equilibrium spectrum, a system of time-dependent kinetic equations for the distribution functions of 
photons, electrons and positrons is solved using the code described in details by \citet{VP09},  extended to include
bremsstrahlung processes. This code is similar to the previous codes of that sort by \citet{coppi92,coppi99}, but includes the 
synchrotron boiler \citep{GGS88,GHS98} and  treats Coulomb thermalization more accurately due to a grid in electron momenta 
extended to low values. 
A code very similar to ours was recently developed by \citet*{BMM08}.

\begin{table*}
\caption{Model parameters and results. \label{tab:results} }
  \begin{center}
    \begin{minipage}[c]{12.5cm}
     \begin{tabular}{@{} rcccccccccc@{}}
\hline
Run & Figure&$M_{\rm BH}/\rm{M_{\odot}}$&$L/L_{\rm Edd}$&$R/R_{\rm S}$&$\Gamma_{\rm inj}$&$\tau$&$\eta_{\rm B}$&$f$&
$\alpha_{\rm [2-10]}$&$k T_{\rm e}$ (keV)\\
\hline
1$^a\!\!\!$ & \ref{fig1_nb}   & $10^7$ & $10^{-2}$            & 10   &  3.0  & 1.0  &  0.1  &  0   &  0.800  &  93\\
2 & \ref{fig1_nb}             & $10^7$ & $10^{-2}$            & 10   &  3.0  & 1.0  &  1.0  &  0   &  0.888  &  95\\
3 & \ref{fig1_nb}             & $10^7$ & $10^{-2}$            & 10   &  3.0  & 1.0  &  10   &  0   &  0.965  &  96\\
\\
4 & \ref{fig2_mass}           & $10$   & $10^{-2}$            & 10   &  3.0  & 1.0  &  0.1  &  0   &  0.779  &  99\\
5 & \ref{fig2_mass}           & $10^3$ & $10^{-2}$            & 10   &  3.0  & 1.0  &  0.1  &  0   &  0.789  &  96\\
6 & \ref{fig2_mass}           & $10^5$ & $10^{-2}$            & 10   &  3.0  & 1.0  &  0.1  &  0   &  0.795  &  94\\
1$^a\!\!\!$ & \ref{fig2_mass} & $10^7$ & $10^{-2}$            & 10   &  3.0  & 1.0  &  0.1  &  0   &  0.800  &  93\\
7 & \ref{fig2_mass}           & $10^9$ & $10^{-2}$            & 10   &  3.0  & 1.0  &  0.1  &  0   &  0.804  &  93\\
\\
8&\ref{fig3_mass}             & $10$   & $10^{-2}$            & 10   &  3.0  & 1.0  &  1.0  &  0   &  0.908  &  94\\
9& \ref{fig3_mass}            & $10^3$ & $10^{-2}$            & 10   &  3.0  & 1.0  &  1.0  &  0   &  0.905  &  93\\
10&\ref{fig3_mass}            & $10^5$ & $10^{-2}$            & 10   &  3.0  & 1.0  &  1.0  &  0   &  0.896  &  94\\
11&\ref{fig3_mass}            & $10^7$ & $10^{-2}$            & 10   &  3.0  & 1.0  &  1.0  &  0   &  0.882  &  95\\
12&\ref{fig3_mass}            & $10^9$ & $10^{-2}$            & 10   &  3.0  & 1.0  &  1.0  &  0   &  0.880  &  96\\
\\
13&\ref{fig4_1e-3Edd}         & $10$   & $10^{-3}$            & 10   &  3.0  & 0.3  &  0.1  &  0   &  0.899  &  122\\
14&\ref{fig4_1e-3Edd}         & $10^3$ & $10^{-3}$            & 10   &  3.0  & 0.3  &  0.1  &  0   &  0.896  &  122\\
15&\ref{fig4_1e-3Edd}         & $10^5$ & $10^{-3}$            & 10   &  3.0  & 0.3  &  0.1  &  0   &  0.895  &  121\\
16&\ref{fig4_1e-3Edd}         & $10^7$ & $10^{-3}$            & 10   &  3.0  & 0.3  &  0.1  &  0   &  0.890  &  121\\
17&\ref{fig4_1e-3Edd}         & $10^9$ & $10^{-3}$            & 10   &  3.0  & 0.3  &  0.1  &  0   &  0.886  &  120\\
\\
18&\ref{fig5_tau}             & $10^7$ & $10^{-2}$            & 10   &  3.0  & 0.1  &  0.1  &  0   &  0.935  &  50--100$^{b}$\\
19&\ref{fig5_tau}             & $10^7$ & $10^{-2}$            & 10   &  3.0  & 0.3  &  0.1  &  0   &  0.889  &  50--100$^{b}$\\
1$^a\!\!\!$ & \ref{fig5_tau}  & $10^7$ & $10^{-2}$            & 10   &  3.0  & 1.0  &  0.1  &  0   &  0.800  &  93\\
20&\ref{fig5_tau}             & $10^7$ & $10^{-2}$            & 10   &  3.0  & 3.0  &  0.1  &  0   &  0.746  &  33\\
21&\ref{fig5_tau}             & $10^7$ & $10^{-2}$            & 10   &  3.0  & 10   &  0.1  &  0   &  0.603  &  7\\
\\
2 & \ref{fig8_Edd}             & $10^7$ & $10^{-2}$            & 10   &  3.0  & 1.0  &  1.0  &  0   &  0.888  &  95\\
22&\ref{fig8_Edd}            & $10^7$ & $10^{-3}$             & 10   &  3.0  & 1.0  &  1.0  &  0   &  0.889  &  102\\
23&\ref{fig8_Edd}            & $10^7$ & $10^{-4}$             & 10   &  3.0  & 1.0  &  1.0  &  0   &  0.890  &  108\\
\\
24&\ref{fig9_inj}             & $10^7$ & $10^{-2}$            & 10   &  2.0  & 1.0  &  0.1  &  0   &  0.915  &  39\\
1$^a\!\!\!$ & \ref{fig9_inj}  & $10^7$ & $10^{-2}$            & 10   &  3.0  & 1.0  &  0.1  &  0   &  0.800  &  93\\
25&\ref{fig9_inj}             & $10^7$ & $10^{-2}$            & 10   &  3.5  & 1.0  &  0.1  &  0   &  0.749  &  105\\
\\
26&\ref{fig10_large_size}       & $10^7$ & $10^{-2}$          & 300  &  3.0  & 1.0  &  1.0  &  0   &  0.886  &  107\\
27$^c\!\!\!$&\ref{fig10_large_size}& $10^7$ & $10^{-2}$       & 300  &  3.0  & 1.0  &  1.0  &  0   &  0.853  &  105\\
\\
1$^a\!\!\!$ & \ref{fig11_disc}& $10^7$ & $10^{-2}$            & 10   &  3.0  & 1.0  &  0.1  &  0   &  0.800  &  93\\
28&\ref{fig11_disc}           & $10^7$ & 0.03                 & 10   &  3.0  & 1.0  &  0.1  & 0.1  &  0.898  &  63\\
29&\ref{fig11_disc}           & $10^7$ & 0.1                  & 10   &  3.0  & 1.0  &  0.1  & 0.3  &  1.030  &  36\\
30&\ref{fig11_disc}           & $10^7$ & 0.3                  & 10   &  3.0  & 1.0  &  0.1  & 1.0  &  1.191  &  16\\
31&\ref{fig11_disc}           & $10^7$ & 1                    & 10   &  3.0  & 1.0  &  0.1  & 3.0  &  1.335  &  4\\
\\
32&\ref{fig13_excess}         & $10^8$ & 1.0                  & 10   &  3.5  & 1.0  &  0.25 & 10   &  1.640  &  3\\
\hline
      \end{tabular}
\begin{center}
  $^{a}$ The fiducial parameter set.
  $^{b}$ The value of the temperature is not well defined as the electron distribution is poorly fitted by the Maxwellian.
  $^{c}$ The spectrum was calculated including bremsstrahlung.
\end{center}
    \end{minipage}
  \end{center}
\end{table*}

\begin{figure}
\centerline{ \epsfig{file=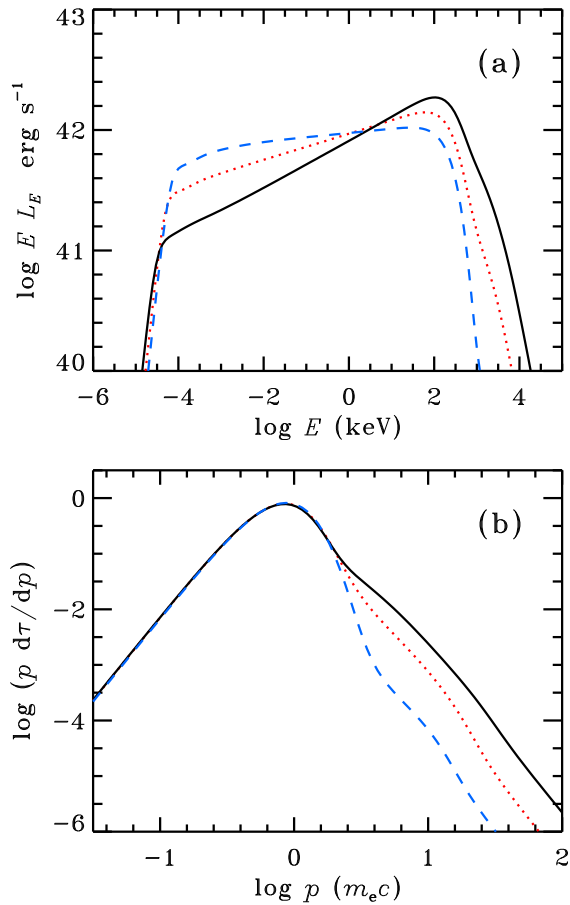, width=7.5cm}}
\caption{Equilibrium spectra (upper panel) and electron distributions (lower panel) for SSC models with the varying 
magnetization parameter: $\eta_{\rm B} = 0.1$ (black solid line), 1.0 (red dotted line), 10 (blue dashed line).
Other parameters are fiducial. Runs 1--3 from Table \ref{tab:results}. 
}
\label{fig1_nb} 
\end{figure}

\section{Results}

In this section we present the detailed investigation of the features of equilibrium photon and electron distributions arising
from variations of
different parameters.
We consider pure SSC models in Section \ref{sec:pure_SSC}.
In Section \ref{sec:brem}, we investigate the role of bremsstrahlung in SSC models. Spectral formation with contribution of the 
accretion disc photons is described in Section \ref{soft_photons}. Specific parameters of each simulation can be found in
Table~\ref{tab:results}.

\subsection{SSC models}
\label{sec:pure_SSC}

Let us first consider pure SSC models putting $f = 0$, i.e. with no contribution from the accretion disc. 
We study how variations of the parameters influence the spectral energy distribution (particularly, the slope $\alpha$ in the
range 2--10 keV) and the electron distribution.
For the fiducial parameters, we take $M_{\rm BH} = 10^7 \rm{M_{\odot}}$, $\Gamma_{\rm inj} = 3.0$, $\tau = 1.0$,
$\eta_{\rm B} = 0.1$, $L/L_{\rm Edd} = 10^{-2}$ (see Fig.~\ref{fig1_nb}, solid line).
The equilibrium electron distribution consists of a Maxwellian part with $k T_{\rm e} = 93$ keV and a power-law tail
with the slope determined by the radiative cooling $s = \Gamma_{\rm inj} + 1 = 4$.
The synchrotron emission coming from thermal electrons is completely self-absorbed and the low-energy part of the spectrum
arises from the synchrotron produced by a non-thermal electron population only (with Lorentz factors $\gamma \gtrsim 30$).
On the other hand, the high-energy Comptonization component, cutting off at $\sim100$ keV, is mostly produced by the Maxwellian
part of the electron distribution.
Above the cut-off energy, there is a high-energy tail resulting from Comptonization on the non-thermal electrons.

\subsubsection{Dependence on the magnetic field}

Let us consider the influence of the strength of magnetic field on spectral formation.
The energy density of typical magnetic fields in the vicinity of black holes is assumed to be of the order of the equipartition
with the radiation field or gas energy density \citep{GRV79}.
In Fig.~\ref{fig1_nb}, the equilibrium spectra and electron distributions are shown for $\eta_{\rm B} = 0.1$, 1.0,
10. When the role of the magnetic field increases, it causes the increasing emission in the low-energy part of the spectrum, where
the main contribution comes from the synchrotron mechanism. 
Subsequently, the electron cooling occurs faster (for the comparison with the re-acceleration time see
equation~\ref{eq:cool/ra}).
This implies the decreasing role of inverse Compton scattering on non-thermal electrons and thus leads to a weaker high-energy
tail in the photon distribution. 
The resulting Comptonized spectrum becomes softer.
As electrons lose more energy in synchrotron emission and cool more efficiently, the non-thermal tail drops and the thermal part
of distribution extends to higher energies.
At the same time, the peak of the thermal distribution is determined by Coulomb processes; thus the equilibrium
temperature is independent of the magnetization.
Although the electron temperatures are almost identical in these simulations, the slopes are different.
This is due to the fact that the slope is determined by both thermal and non-thermal electron populations, with a harder
spectral index in the case of more powerful electron tail.

One can notice in Fig.~\ref{fig1_nb}(b), that the normalization of the electron tail at $\gamma < 10$ does not scale inversely
proportional to magnetization.
The competing processes determining the normalization (and shape) of the high-energy tail are Compton and
synchrotron cooling.
Their relative role in tail formation is determined by the magnetization parameter (see equation ~\ref{eq:cools/cs}).
With the increasing role of the magnetic field, the cooling becomes dominated by synchrotron
processes, and from this moment the normalization scales inversely with $\eta_{\rm B}$.
Also at high Lorenz factors, Compton scattering plays a less important role, because cooling occurs partially in Klein-Nishina
regime.

\begin{figure}
\centerline{ \epsfig{file=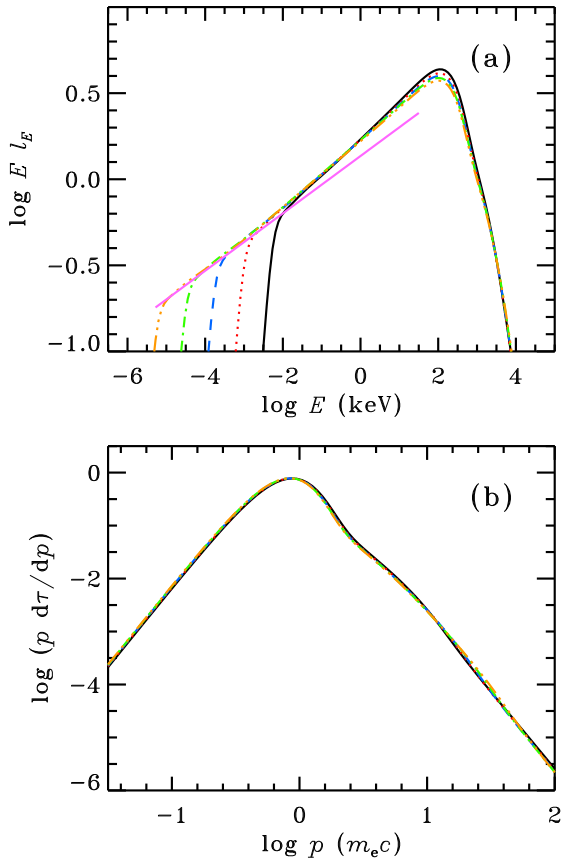,width=7.5cm}}
\caption{Equilibrium spectra (differential compactness, upper panel) 
and electron distributions (lower panel) for SSC models with varying masses of the black hole: 
$M=10$ (black solid line), $10^3$ (red dotted line), $10^5$ (blue dashed line), $10^7\rm{M_{\odot}}$ (green dot-dashed line) and
$10^9\rm{M_{\odot}}$ (yellow three-dot-dashed line). 
Other parameters are fiducial.
The broadest spectrum corresponds to the highest mass. 
The magenta line corresponds to the power-law of the energy index $\alpha=5/6$.
Runs 1, 4--7 from Table \ref{tab:results}. }
\label{fig2_mass} 
\end{figure}

\subsubsection{Dependence on the mass}

Variations in the mass correspond to variations of two parameters simultaneously: the Schwarzschild radius and the Eddington
luminosity. 
If we fix the Eddington ratio and the size (in Schwarzschild radii), then the compactness parameter becomes independent of the
mass.
The spectral energy distribution $l_E$ is not expected to change much. 
However, the temperature of the accretion disc photons as well as the magnetic field strength (and characteristic synchrotron
frequencies) do  depends on mass. 

If the seed photons for Comptonization come from an accretion disc, the size and disc temperature scale as $R \propto M_{\rm BH}$
and $T_{\rm d} \propto M_{\rm BH}^{-1/4}$, respectively \citep*[see, e.g.][]{FKR02}.
For the black holes with higher masses and  the same compactness, the Comptonized spectrum becomes broader 
and the X-ray spectra of Seyferts are expected to be softer than those of GBHs at the same ratio $f$ \citep{ZLG03} for
$\alpha<1$.

\begin{figure}
\centerline{ \epsfig{file=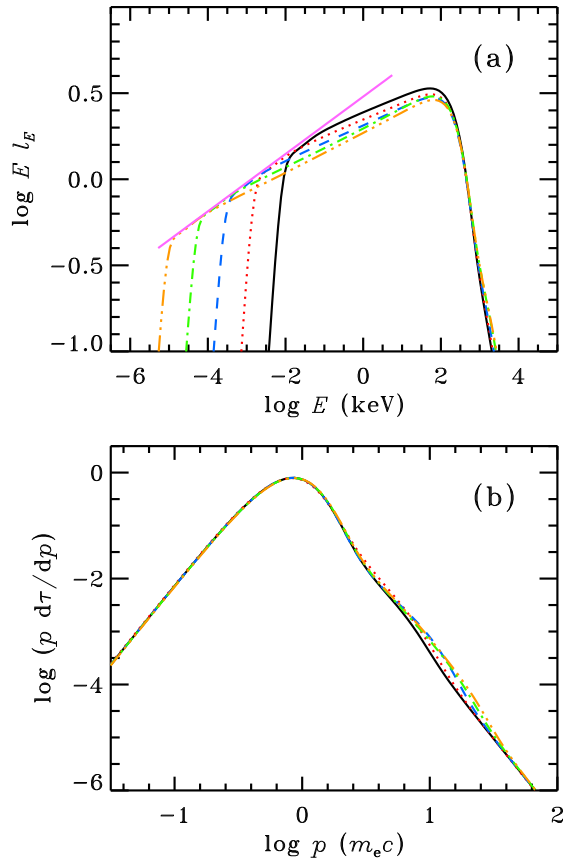,width=7.5cm}}
\caption{Same as in Fig.~\ref{fig2_mass}(a), but for $\eta_{\rm B} = 1.0$. 
Runs 8--12 from Table \ref{tab:results}. 
}
\label{fig3_mass} 
\end{figure}

\begin{figure}
\centerline{ \epsfig{file=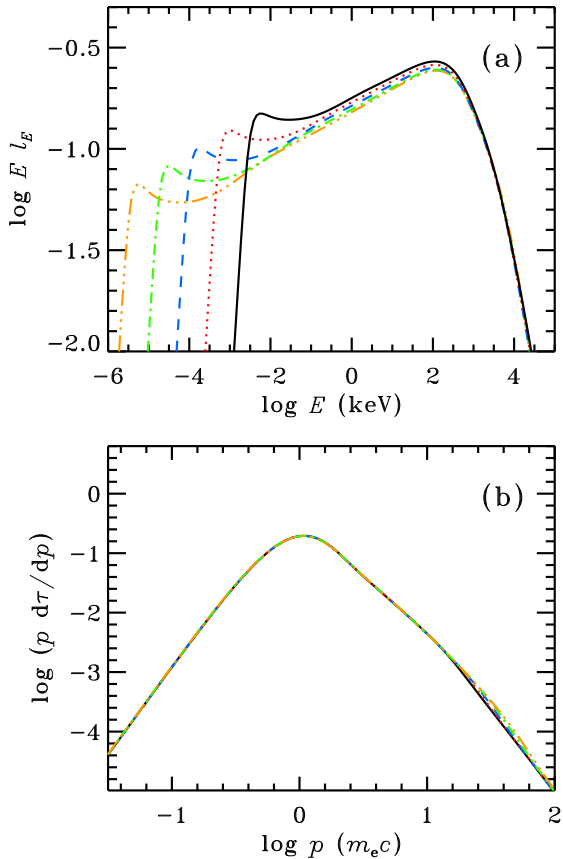,width=7.5cm}}
\caption{Same as in Fig.~\ref{fig2_mass}, but for $L/L_{\rm Edd} = 10^{-3}$ and $\tau = 0.3$.
Runs 13--17 from Table \ref{tab:results}.  }
\label{fig4_1e-3Edd} 
\end{figure}

The situation is more complicated for the SSC mechanism.
The low-energy part of the spectrum is determined by synchrotron radiation which is strongly self-absorbed up to some critical
frequency, the so-called turnover frequency $\nu_{\rm t}$. 
For hybrid electron distribution, containing Maxwellian and power-law parts, there are two corresponding turnover frequencies:
thermal
$\nu_{\rm t}^{\rm th} \propto B^{0.91}$ \citep[e.g.][]{Mah97, WZ00} and power-law \citep[e.g.][]{RL79, WZ01}
\begin{equation} \label{eq:nutB}
\nu_{\rm t}^{\rm  pl} \propto B^{\frac{s+2}{s+4}} \propto R^{-\frac{s+2}{2s+8}}  \propto M_{\rm BH}^{-\frac{s+2}{2s+8}}, 
\end{equation}
where we used equation (\ref{eq:magfield}), obtaining the last scaling.
Here $s = \Gamma_{\rm inj}+1$, if the cooling of particles emitting around the turnover frequency is dominated by
synchrotron radiation or Compton scattering and $s = \Gamma_{\rm inj} - 1$, if Coulomb collisions dominate (see
equation~\ref{eq:elslope}).
In our simulations, the synchrotron radiation coming from the Maxwellian part of the electron distribution is completely
self-absorbed and only non-thermal electrons are responsible for the synchrotron radiation coming from the region.
Using equation (\ref{eq:magfield}) we then deduce (for $\Gamma_{\rm inj} = 3.0$)
\begin{equation}
 \nu_{\rm t} \propto \left( \frac{\eta_{\rm B} l}{R} \right)^{0.33 \div 0.38}.
\end{equation}
The lower and upper limits of the power-law index correspond to Coulomb and radiative regimes, respectively. 
Moreover, for the pure SSC simulations, the synchrotron-emitting particles are in the radiative cooling
regime (i.e. the cooling by Coulomb is negligible). 
The turnover frequency is generally smaller for more massive objects (see Figs~\ref{fig2_mass}--\ref{fig4_1e-3Edd}) and lower
magnetizations.
The latter effect can be seen in Fig.~\ref{fig1_nb}; however, the dependence on $\eta_{\rm B}$ is weak since the
effects of the increasing magnetic field and decreasing normalization of the tail of the electron distribution due to stronger
cooling nearly cancel each other out.
We find that for different values of magnetization parameter, 
the slope of the thermal Comptonization component can become harder or softer with increasing mass 
(Figs.~\ref{fig2_mass} and \ref{fig3_mass}),
depending on its initial value and the slope of the non-thermal electron distribution. 
The electron distribution is independent of the mass due to the fact that cooling and heating (re-acceleration) times scale with
the mass similarly, thus their ratio is independent of the mass (see equations \ref{eq:cool/ra} and \ref{eq:Coul/ra}).

\begin{figure}
\centerline{ \epsfig{file=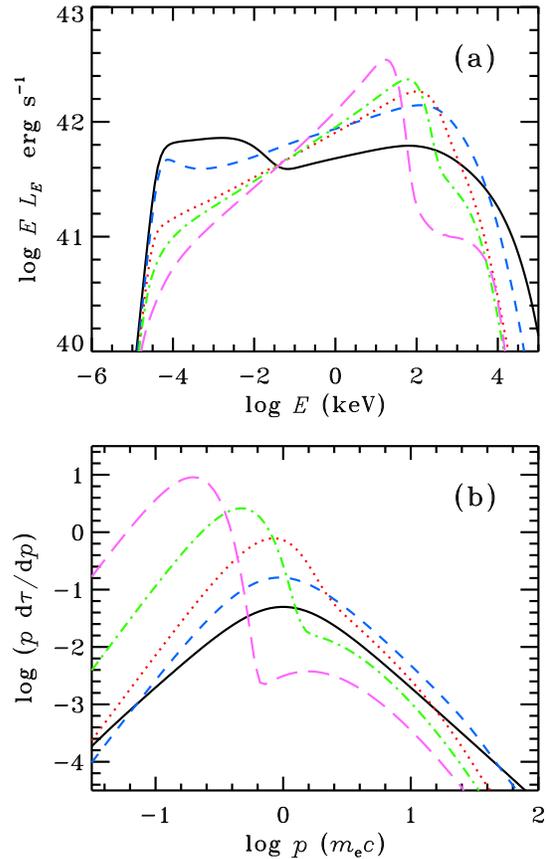,width=7.5cm}}
\caption{Equilibrium spectra (upper panel) and electron distributions (lower panel) for SSC models with varying optical depths
$\tau$ = 0.1, 0.3, 1, 3, 10. The hardest spectrum and the coldest electron distribution correspond to the highest $\tau$. 
Other parameters are fiducial.
Runs 1, 18--21 from Table \ref{tab:results}. 
}
\label{fig5_tau} 
\end{figure}

The synchrotron luminosity at $\nu_{\rm t}$ can be estimated by remembering that the intensity at the turnover frequency is just
a source function of the power-law electrons and that the relativistic electron of the Lorentz factor $\gamma$ emits most of the
synchrotron power at frequency $\nu \simeq 2 \gamma^2 \nu_{\rm B}$ (where $\displaystyle \nu_{\rm B} =eB/2\pi m_{\rm e} c$ is the
cyclotron frequency). Then, the compactness of the synchrotron photon spectrum at  $\nu_{\rm t}$ scales as \citep[][]{RL79}
\begin{equation}\label{eq:ele_scaling}
 E \, l_{\rm E, synch} \propto \frac{R \nu_{\rm t}^{7/2}}{\sqrt{\nu_{\rm B}}} \propto  B^{\frac{s-3}{s+4}} 
 \propto  R^{-\frac{s-3}{2s+8}} \propto M_{\rm BH}^{-\frac{s-3}{2s+8}}. 
\end{equation}
Eliminating the mass using equation (\ref{eq:nutB}), we get the dependence of differential compactness (at a turnover frequency)
on the photon energy
\begin{equation} \label{eq:ssa_sp_scaling}
 E \, l_{\rm E, synch} \propto E^{\frac{s-3}{s+2}}.
\end{equation}
This corresponds to a power-law of the energy index $\alpha=5/(s+2)$. 
In Figs~\ref{fig2_mass} and \ref{fig3_mass}, the magenta line corresponds to $s = \Gamma_{\rm inj} + 1 = 4$, 
giving  $E \, l_{\rm E, synch} \propto E^{1/6}$ (i.e. $\alpha=5/6$). 
For an initial spectral index $\alpha$, increasing the mass of the compact object leads to softer SSC spectra if 
$s < 5/\alpha - 2$.
For the canonical value $\alpha \approx 0.7$ in the hard-state spectra of GBHs we get $s < 5$ (or equivalently
$\Gamma_{\rm inj} < 4$). 
For a steep spectral slope to start with at 10 $\rm{M_{\odot}}$ (for instance, the slope as in Fig.~\ref{fig3_mass}), we get a
more stringent constraint then $s< 5/0.908 - 2 \approx 3.5$ and  $\Gamma_{\rm inj} < 2.5$. 
Thus, a hard electron injection function is required to get the systematically softer spectra in Seyferts in terms of the SSC
mechanism, keeping all the parameters (except $M_{\rm BH}$) constant.
 
We emphasize here that whether the spectrum hardens or softens with the increasing mass depends both on the initial 
(for $M_{\rm BH} = 10 \rm{M_{\odot}}$) slope and the scaling of the seed photon compactness.
The power-law slope of the Comptonization component will tend asymptotically to the value $\alpha=5/(s+2)$ (thus, hardening or
steepening) with the increasing mass of the object independently of other parameters, leading to a stability of the spectrum. 
In models with the seed photons coming from the accretion disc (with $f$ = const) this asymptotic slope is $\alpha=1$.

In Fig.~\ref{fig4_1e-3Edd} we  plot spectra and electron distributions obtained for the same parameters as in
Fig.~\ref{fig2_mass}, but for lower $\tau$ and $L/L_{\rm Edd}$. 
The tail in the electron distribution becomes relatively stronger resulting in more luminous synchrotron emission. This softens
the Comptonized part of the spectrum (compared to the cases in Fig.~\ref{fig2_mass}). 
The slopes of electron distributions between momenta $p=$ 1 and 10 resemble the injection function, implying inefficient
cooling at these energies. Below the peak, the Coulomb collisions dominate the re-acceleration, leading to a formation of the
Maxwellian distribution, and for $p > 10$, radiative cooling softens the electron distribution.

\begin{figure}
\centerline{ \epsfig{file=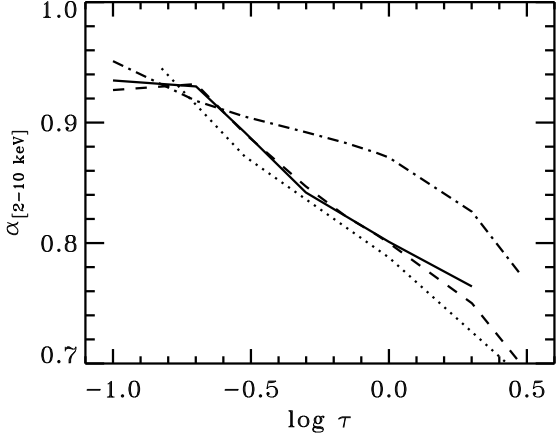,width=7.5cm}}
\caption{
Dependence of the spectral slope on the optical depth for various sets of parameters 
($M_{\rm BH}/\rm{M_{\odot}}$, $L/L_{\rm Edd}$, $\Gamma_{\rm inj}$)=
($10^7$, $10^{-2}$, 3.0) (solid line), ($10^7$, $10^{-3}$, 3.0) (dashed line), ($10^7$, $10^{-2}$, 2.0) (dot-dashed line). 
The corresponding dependence for GBHs with parameters ($10$, $10^{-2}$, 3.0) is shown with the dotted line. Other parameters are
fiducial.
}
\label{fig6_tau_compare} 
\end{figure}

\begin{figure}
\centerline{ \epsfig{file=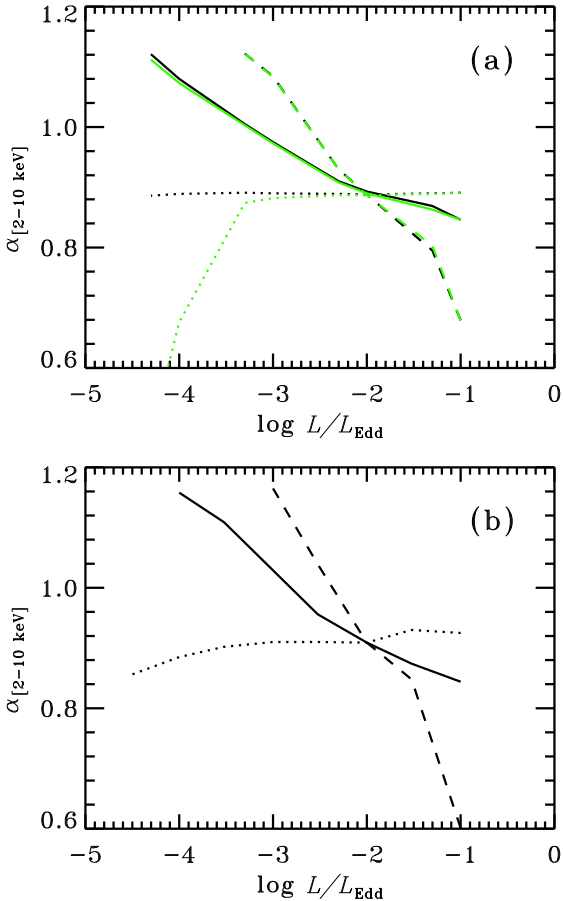,width=7.5cm}}
\caption{
Dependence of the spectral slope on the Eddington ratio for for pure SSC models. 
(a) SMBH ($M_{\rm BH}=10^7 \rm{M_{\odot}}$, $\eta_{\rm B} = 1.0$) and 
(b) GBH ($M_{\rm BH}=10 \rm{M_{\odot}}$, $\eta_{\rm B} = 1.0$). 
Different dependencies of optical depth on the luminosity are shown: $\tau \propto L^{1/2}$ (solid line), 
$\tau \propto L$ (dashed line), $\tau=$constant (dotted line). The curves intersect at $L = 10^{-2} L_{\rm Edd}$ where
$\tau = 1.0$.
The light (green) curves correspond to all radiation processes taken into account, while for the black curves the bremsstrahlung
is neglected.
}
\label{fig7_Edd_compare} 
\end{figure}

\subsubsection{Dependence on the optical depth}
\label{sec:opt_depth}

Now let us consider the role of the optical depth. Again, we take the fiducial parameters and let $\tau$ vary. 
As the optical depth grows, the electrons have more time to cool before being re-accelerated, leading to a drop in the
electron temperature (Fig.~\ref{fig5_tau}b).
Lower temperature and higher optical depth lead to efficient thermalization by Coulomb collisions, driving a larger fraction of
particles to a Maxwellian distribution. 
As the electron temperature decreases, the peak of the thermally Comptonized spectrum also moves to lower energies (see
Fig.~\ref{fig5_tau}a). At the same time, the number of particles in the non-thermal tail of the electron distribution decreases,
leading to weaker partially self-absorbed synchrotron emission.
The source therefore becomes photon-starved and the spectrum hardens.

At low optical depths the particles do not have time to cool between re-accelerations and the electron distribution resembles
the injection function with a hard tail.
The synchrotron emission is therefore strong and dominates the emission in the 0.1 -- 10 eV range for $\tau = 0.1$.

The effect of the hardening of the spectrum with increasing optical depth was found for both GBHs and SMBHs 
irrespective of the values of the other parameters (see Fig.~\ref{fig6_tau_compare}), 
suggesting that it has the strongest influence on the slope.

\begin{figure}
\centerline{ \epsfig{file=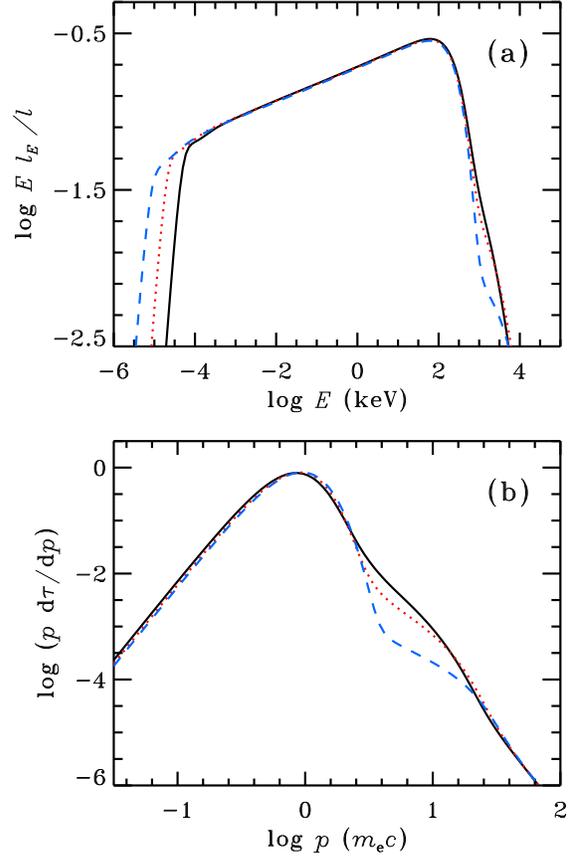,width=7.5cm}}
\caption{Equilibrium spectra (normalized to the total compactness, upper panel) and electron distributions (lower
panel) for pure SSC models with varying Eddington ratio: $L/L_{\rm Edd} = 10^{-2}$ (black solid line), $10^{-3}$ (red dotted
line), $10^{-4}$ (blue dashed line).
The optical depth is constant in these simulations. $\eta_{\rm B} = 1.0$, other parameters are fiducial. 
Runs 2, 22 and 23 from Table \ref{tab:results}. }
\label{fig8_Edd} 
\end{figure}

\begin{figure}
\centerline{ \epsfig{file=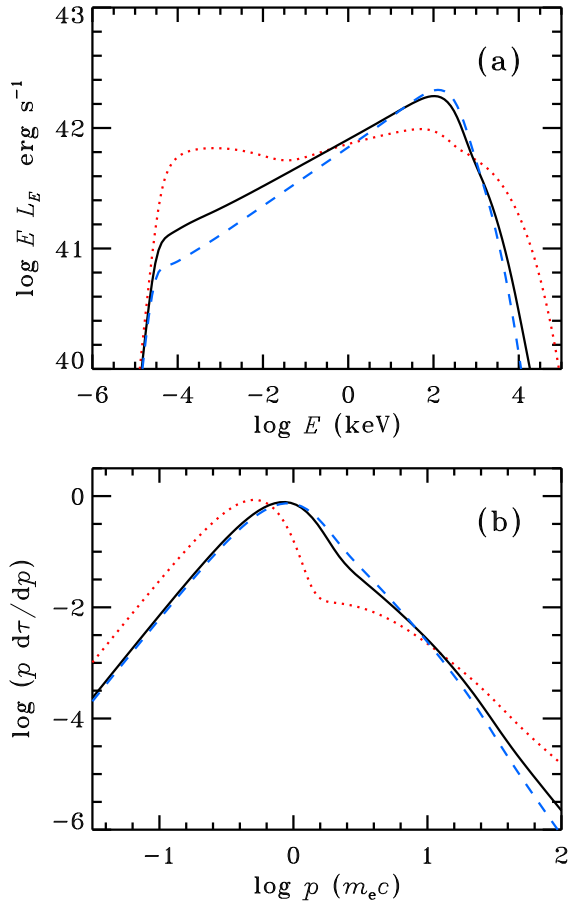,width=7.5cm}}
\caption{
Equilibrium spectra (upper panel) and electron distributions (lower panel) for pure SSC models with varying 
injection slopes: $\Gamma_{\rm inj} = 2.0$ (red dotted line), $3.0$ (black solid line), $3.5$ (blue dashed line).
Other parameters  are fiducial. Runs 1, 24 and 25 from Table \ref{tab:results}. }
\label{fig9_inj} 
\end{figure}

\subsubsection{Dependence on the Eddington ratio}

The change in the Eddington ratio is closely connected to the change of the accretion rate, which, in turn, affects
the optical depth. The latter is rather obvious: if the accretion time is constant, the higher is the matter supply, the higher is
the density and therefore the optical depth. We assume that $\tau \propto \dot{M}$.
For different types of accretion modes (corresponding to efficient and inefficient flows), the observed X-ray luminosity
depends on the accretion rate in a different manner.
To investigate spectral formation at various accretion rates, we consider a simple power-law scaling 
\begin{equation}\label{eq:optdepth_lum}
 \tau = \tau_0 \left( \frac{L}{L_{\rm Edd}} \right)^{\theta},
\end{equation}
where $\tau_0 = 1.0$ and $\theta = 0.5$, which corresponds to radiatively inefficient accretion ($\tau \propto \dot{M}$ and
$L \propto \dot{M}^2$), $\theta = 1.0$ (corresponding to radiatively efficient accretion) and $\theta = 0$, i.e. 
constant optical depth (ad hoc).

The results of simulations are shown in Figs~\ref{fig7_Edd_compare} and \ref{fig8_Edd}. The spectral index is almost independent
of the Eddington ratio, if the optical depth remains constant and  bremsstrahlung is not accounted for. 
The turnover frequency is smaller for lower Eddington ratios: the fraction of the magnetic energy ($\eta_{\rm B}$) remains
constant; therefore, the magnetic field itself decreases as $B \propto (L/L_{\rm Edd})^{1/2}$, leading to the reduction of
$\nu_{\rm t}$.
The spectral slopes in the 2--10 keV range are determined by Comptonization on thermal electrons. Our simulations show that for
the case of constant optical depth, the Maxwellian part of the distribution remains almost unchanged. 
The Comptonized slope can be approximated as \citep{B99ASP}:
\begin{equation} \label{eq:spectral_index}
 \alpha \simeq \frac 94 y^{-2/9} - 1,
\end{equation}
where $y = 4 (\Theta_{\rm e} + 4\Theta_{\rm e}^2) \tau (\tau+1)$ is the Compton parameter and $\Theta_{\rm e} = kT_{\rm e}/m c^2$
is the dimensionless temperature of electron distribution. 
Hence, for constant optical depth, equal temperatures of the Maxwellian distribution give the same 
slope of the Comptonization spectrum  (see Fig.~\ref{fig8_Edd}).
When the optical depth changes together with the Eddington ratio, the electron temperature is no longer conserved, and the
spectral slope changes too.
At sufficiently low luminosities and relatively high optical depths the X-ray spectrum is no longer dominated by Comptonization,
but instead determined by bremsstrahlung.

Because the injection and cooling rates scale the same way with the Eddington ratio, the electron distribution at
energies producing synchrotron emission above the turnover frequency
\begin{equation}
 \gamma  \gtrsim 
 \gamma_{\rm t} \simeq \sqrt{ \frac {\nu_{\rm t}}{2 \nu_{\rm B}}} \approx 4.2 \times 10^{-4} \sqrt{ \frac {\nu_{\rm t}}{B} }
 \approx 30 
\end{equation}
has a power-law shape with $s=\Gamma_{\rm inj}+1$ and does not change in the cases plotted in Fig.~\ref{fig8_Edd}.
Using equations~(\ref{eq:nutB}) and (\ref{eq:ele_scaling}), one gets the dependence of $E l_{\rm E}/l$ (the spectral
distribution scaled to the total compactness) on the turnover frequency for the constant $R$:
\begin{equation}
 \frac {E l_{\rm E}}{l} \propto \frac {R \nu_{\rm t}^{7/2}}{l \sqrt{\nu_{\rm B}}} \propto E^{\frac {s-3}{s+2}},
\end{equation}
giving $\alpha = 5/(s+2) = 5/6$, which happens to be very close to the X-ray spectral index 0.89.
Thus, the spectral slope in this case is conserved.

The three electron distributions differ much in the range of Lorentz factors between $\gamma\sim$ 3 and 20.
The effect is due to the increasing role of Coulomb cooling for lower Eddington ratios (comparing to radiative cooling, see
equation~\ref{eq:Coul/cs}). For the adopted parameters and $\gamma \approx 10$, the ratio of radiative (Compton) and Coulomb
time-scales is
\begin{equation}
 \frac {t_{\rm cool, Coul}}{t_{\rm cool, cs}} \approx 2.8\, l_{\rm inj}.
\end{equation}
Thus, for $L/L_{\rm Edd} = 10^{-2}$ ($l_{\rm inj} \approx 10$), the Compton cooling dominates, but for 
$L/L_{\rm Edd} = 10^{-4}$ (with $l_{\rm inj} \approx 0.1$), the slope of the electron tail at $\gamma$ between 3 and 20 is
predominantly determined by Coulomb collisions and $s \approx \Gamma_{\rm inj} - 1 = 2$ (see equation~\ref{eq:elslope}).
Coulomb collisions also cause the heating of thermal electrons to slightly higher temperatures.

\subsubsection{Dependence on the injection slope}

The exact mechanism responsible for the particle acceleration in the inner parts of accretion flow is poorly understood. 
It can be related to weak shocks within the accretion flow or reconnection of the magnetic field. 
The slope of the injected electrons is strongly model-dependent. 
On the observational side, the slope in GBHs can be estimated from the MeV tails \citep{McConnell02}, 
but for AGNs the observations are not yet very constraining \citep{Gon96,Johns97}. 

To investigate the dependence of the results on the injection slope, we consider three cases with $\Gamma_{\rm inj} =$ 2.0, 3.0
and 3.5.
The variable injection slope influences the spectrum mainly through its strong effect on the strength of partially self-absorbed
synchrotron radiation produced by non-thermal electrons.
Generally, the synchrotron spectrum from the hybrid electron distribution can be divided into four parts \citep{WZ01}: at low
energies, emission and absorption are dominated by thermal electrons (Rayleigh-Jeans spectrum), at slightly higher energies, the
emission from non-thermal electrons becomes dominant, but the absorption is still determined by the thermal population.
Then the non-thermal absorption becomes more important, and the spectrum is proportional to the source function of non-thermal
electrons.
Finally, at high energies, the non-thermal emission becomes optically thin with the spectral index $\alpha = (s-1)/2$.

The cases $\Gamma_{\rm inj} = 2.0$ and 3.5 are qualitatively different, as in the former case 
the number of electrons emitting above the turnover frequency is large and the synchrotron spectrum is flat (with $\alpha=1$
corresponding to $s=3$, see Fig.~\ref{fig9_inj}). This leads to a large amount 
of soft photons available for Comptonization, strong cooling of the electrons and a softer Comptonized spectrum. 
For high values of $\Gamma_{\rm inj}$, the number of electrons emitting above the turnover frequency is reduced together 
with the synchrotron luminosity causing higher electron temperature and a harder Comptonized spectrum.

\begin{figure}
\centerline{ \epsfig{file=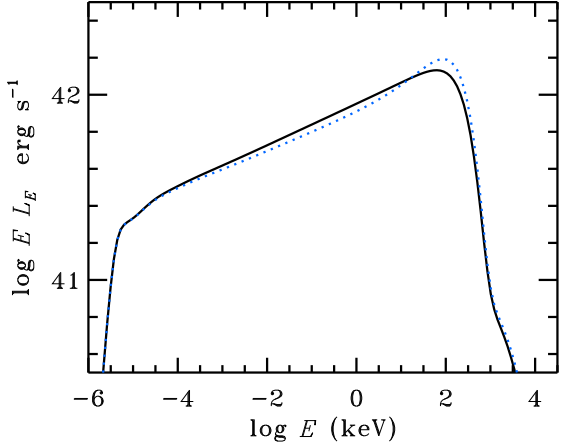,width=7.5cm}}
\caption{Spectra of a SMBH, where the size of the active region is assumed to be $R = 300 R_{\rm S}$. 
Runs 26 and 27 from Table~\ref{tab:results}. Spectra are calculated without accounting for bremsstrahlung processes (black
solid line) and including those (blue dotted line).}
\label{fig10_large_size} 
\end{figure}

\subsection{The role of bremsstrahlung in SSC models}
\label{sec:brem}

We have discussed the spectral formation under the joint action of synchrotron and Compton processes.
Let us now consider how the free-free emission affects the resulting spectra.
Major contribution comes from electron-electron and electron-proton collisions (the amount of positrons is negligibly small in
our simulations; thus, the number of protons is approximately equal to number of electrons).
The bremsstrahlung emission coefficients depend on the square of the number density of electrons, while the Compton emissivity
depends on the product of the number density of particles and photons. 
Generally one would expect the former processes to play dominant role in the medium with high particle and low photon number
densities.
The relevant scalings are $n_{\rm e} \propto \tau/R$, and
$n_{\rm ph} \propto L t_{\rm esc}/R^3 \propto L/R^2$. Then, the bremsstrahlung emission coefficient scales as $j_{\rm br} \propto
\tau^2/R^2$, and
Compton
scattering emission as $j_{\rm cs} \propto L \tau/R^3$.
The relative importance of bremsstrahlung emission is thus characterized by the ratio
\begin{equation}\label{eq:brem_imp}
 j_{\rm br}/j_{\rm cs} \propto \tau R/L \propto \tau/l.
\end{equation}
The effect of bremsstrahlung on the spectral slope for different dependencies of $\tau$ on $L/L_{\rm Edd}$ can be seen in
Fig.~\ref{fig7_Edd_compare}.
From equation~(\ref{eq:brem_imp}) we see that for a given Eddington ratio (and size), its importance primarily depends on the
optical thickness of the
source. In the cases considered in Fig.~\ref{fig7_Edd_compare}, the free-free radiation only affects the slope in the case when
$\tau$ is kept constant (i.e. large). 
Thus, bremsstrahlung processes can be energetically important for the system of larger size (and small compactness). 
This can correspond to some type of radiatively inefficient accretion \citep{NMQ98}. 
An example of the resulting spectrum is shown in Fig.~\ref{fig10_large_size}.
If the optical depth decreases together with the Eddington ratio, bremsstrahlung emission remains negligible.
 
However, even if bremsstrahlung is energetically not significant, it still can produce a hardening of the otherwise 
power-law-like spectrum above a few keV making the overall spectrum concave.
Such a spectral complexity might have been a reason for existence of soft excess in a GBH Cyg X-1 \citep{IPG05}.
In Seyferts, the data are not of the same quality to detect such a hardening. 

\begin{figure}
\centerline{ \epsfig{file=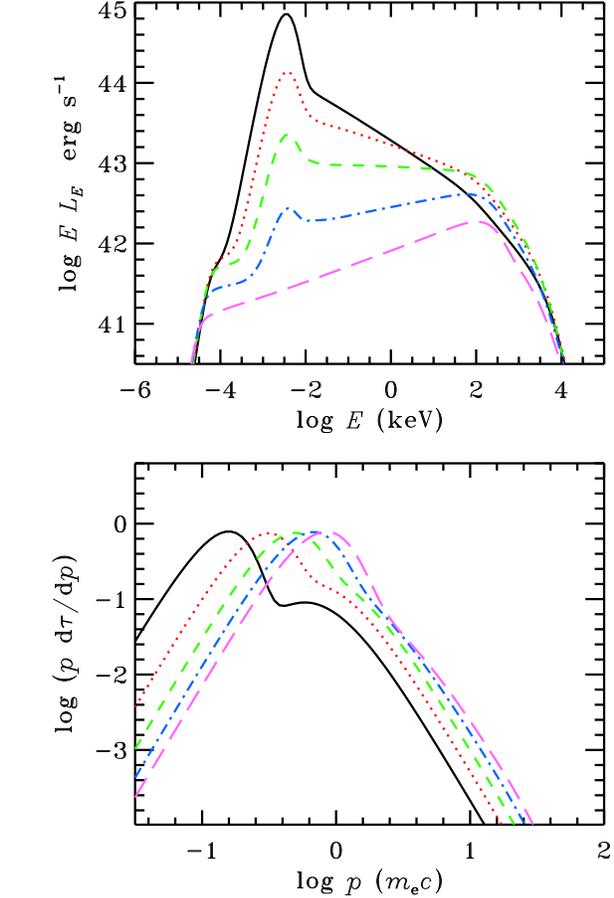,width=7.5cm}}
\caption{Equilibrium spectra and electron distributions for different Eddington ratios  
(from the top to bottom) $L/L_{\rm Edd}$ = 0.01, 0.03, 0.1, 0.3 and 1, and varying disc contribution $f$ = 0, 0.1, 0.3, 1 and 3. 
Runs 1, 28--31 from Table ~\ref{tab:results}. The disc temperature is assumed to be $T_{\rm d} = 10^{4}$ K. }
\label{fig11_disc} 
\end{figure}

\subsection{Influence of the accretion disc and slope -- luminosity correlation}
\label{soft_photons}

So far have considered pure SSC models, where the only source of soft seed photons is synchrotron radiation. This regime can be
realized at sufficiently low accretion rates, when the disc is far away from the region in which the majority of energy is
liberated. 
For black holes in X-ray binaries, there is a strong evidence that the disc inner radius changes with the accretion rate.
The spectral slope correlates with the amplitude of reflection, the width of the iron line, and the characteristic frequencies of
variability (\citealt{ZLS99}; \citealt*{GCR99, RGC01}; \citealt{IPG05}).
Similar correlations are found in Seyfert galaxies in individual sources \citep{MBZ98,Nandra00}
as well as in the set of AGNs \citep{ZLS99,LZ01, Matt01,PPM02,ZLG03}. 

The spectral transitions in GBHs are characterized by the dramatic change in the spectral shape with corresponding changes in the
electron distribution. In the hard state, the electrons are mostly thermal probably only with a weak tail which is responsible for
the production of MeV photons \citep{McConnell02}. In the soft state, the electrons have mostly non-thermal, power-law-like
distribution producing long tail extending up to 10 MeV \citep{PC98,P98}.
The strong variation in the electron distribution can be caused mostly by the increasing Compton cooling  when the disc moves in
as the accretion rate increases \citep{VP08, PV09,MB09}.
Here we repeat simulations presented in aforementioned papers, but for the parameters typical for SMBHs. 

Once the disc moves sufficiently close to the X-ray emitting region, the seed photons from the disc start dominating over the
internally produced synchrotron photons. 
We simulate this process by injecting additional blackbody photons to the simulation volume.  
We parametrize the importance of these photons by the relative fraction to the luminosity dissipated in the active region 
$f \equiv L_{\rm disc}/ L_{\rm inj}$. These seed photons are injected homogeneously into the region. We expect that $f$ is a
strong function
of the cold inner disc radius, 
which in turn depends on the accretion rate and total luminosity. We assume a power-law dependence 
\begin{equation}\label{eq:discfrac}
 f = f_0 \left( \frac{L}{L_{\rm Edd}} \right)^{\beta},
\end{equation}
where $\beta$ and $f_0$ are the parameters to be determined from the data or theory. 
The results of simulations for $\beta=1$ and $f_0=3$ are shown in Fig.~\ref{fig11_disc}. 
As the fraction $f$ increases, the disc quickly replaces synchrotron radiation as a dominant source of seed photons for Compton
upscattering. Strong soft radiation field implies faster cooling, leading to a decrease of the electron temperature and a more
pronounced high-energy tail.
This, in turn, leads to the decline of the Compton parameter associated with the thermal population and the spectrum softens
(see equation~\ref{eq:spectral_index}). The role of thermal Comptonization in the formation of the high-energy spectrum becomes
progressively weaker and is taken over by Compton scattering on the non-thermal population of electrons. For high $f$, the
spectrum is dominated by the disc blackbody, whereas the soft non-thermal tail extends to several MeV without a cut-off. 
The shape of the spectrum above $\sim100$ keV strongly depends on the assumed maximum electron Lorentz factor 
$\gamma_{\rm he}$.

\begin{figure*}
\centerline{ \epsfig{file=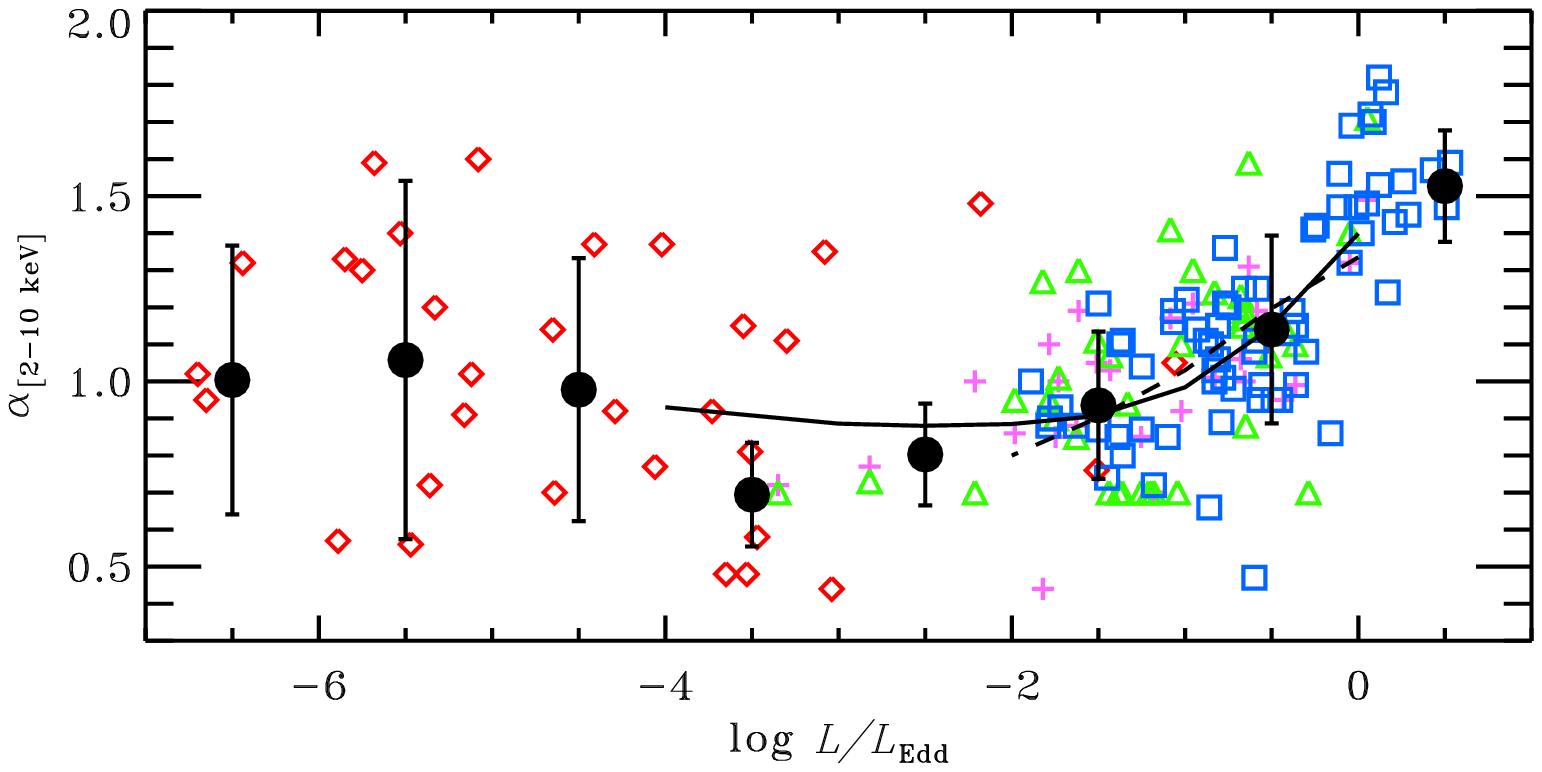,width=14cm}}
\caption{
Dependence of the spectral slope on the Eddington ratio. The red diamonds mark LINERs \citep[compilation from][]{SDO05,GMM09}.
Other symbols correspond to the data on Seyfert galaxies and quasars: magenta crosses \citep{ZhouZhao10}, blue squares
\citep{ZZhang10} and green triangles \citep{VF09}, offset up by 0.2 to correct for reflection.
The black filled circles show the averaged data (binned into intervals of one decade in $L/L_{\rm Edd}$) with the corresponding
dispersion.
The black solid line shows the result of our simulations with parameters 
$f_0 = 2$ and $\beta = 1$ from equation (\ref{eq:discfrac}) and $\tau_0 = 1.0$ and $\theta = 0.5$ from equation (\ref{eq:optdepth_lum}). 
The dashed line corresponds to  $f_0 = 3$ and $\beta = 1$  and the constant optical depth $\tau = 1.0$ 
(corresponding spectra and electron distributions are shown in Fig.~\ref{fig11_disc}).
}\label{fig12_corr} 
\end{figure*}

\section{Observed spectral properties of supermassive black holes}

Among first surveys on X-ray spectral indices of AGNs, \citet{NanPoun94} have noticed that majority of Seyfert 1
galaxies have spectral energy index $\alpha = 0.95$ with small dispersion, $\sigma = 0.15$.
However, addition of NLSy1s broadened the range of spectral indices \citep*{BME97}.
After that, the energy indices $\alpha \sim 1.0$ were also found for $z \lesssim 6$ type I radio-quiet quasars 
\citep[e.g.,][]{PROB04,SBN06}.
Lately, it became clear that dispersion depends on the sample: the more objects are considered, the wider is the spread 
\citep[see, e.g., Fig.~8 in][]{BAG10}.
Also recent studies revealed tight correlation between the spectral slope and the luminosity of the object in Eddington units
(e.g., \citealt*{WWM04}; \citealt{SBN06,VF07,MDS08,Sob09,VF09,ZhouZhao10}).
In contrast, the low-luminosity AGNs (e.g. LINERs), tend to have rather stable spectral slopes with the average $\alpha\sim1$. 

A standard view is that at low mass accretion rates the vicinity of the black hole is occupied by 
a hot radiatively inefficient accretion flow (\citealt{NMQ98}; \citealt*{YQN03,Yuan10}). 
In the case of very large accretion rates typical for NLS1s and quasars, the standard cold accretion disc 
extends all the way down to the last stable orbit around the central black hole with the hard X-rays being probably 
produced in the accretion disc corona. 
In the intermediate case represented by Seyfert galaxies, the cold disc is probably truncated and the 
X-rays are produced in the inner hot flow.  
Let us now apply our model to the observed properties of AGNs at different accretion rates. 

We assume the dependence $f(L)$ given by equation~(\ref{eq:discfrac}) together with $\tau(L)$ given by
equation~(\ref{eq:optdepth_lum}).
The coefficients and the indices can be obtained from the direct comparison with the data.
Fig.~\ref{fig12_corr} (solid line) shows the dependence of the spectral slope on the Eddington ratio for a model with
$\tau = \left(L/L_{\rm Edd}\right)^{1/2}$ and $f = 2 \left(L/L_{\rm Edd}\right)$.
The model describes well the observed slopes for a very broad range of luminosities, as we show below.
The self-absorption frequency is sufficiently low for objects with high masses and the synchrotron emission falls in the
infrared range.
This model predicts an intimate relation between the infrared/optical and X-ray variability in low-luminosity objects.

\subsection{LINERs and stability of the spectral slope}
\label{sec:app_LINERs}

Low-ionization nuclear emission-line regions (LINERs) are defined by narrow optical emission lines of low ionization 
\citep*{Hec80}.
Some fraction of LINERs are believed to host an AGN, with mass accretion rates far below Eddington.
The X-ray spectrum can well be fitted with a single power-law of $\alpha \sim 1.0 - 1.2$ \citep{GMM09}, which shows very little
intrinsic absorption and weak (or absent) signatures of Fe K$\alpha$ emission or Compton reflection 
\citep[][and references therein]{Ho08}.
Thus, the data suggest no accretion disc in the vicinity of the central black hole, 
making LINERs appropriate candidates for pure SSC mechanism.
We have taken the spectral indices from the sample of \citet{GMM09} and the corresponding Eddington ratios from \citet{SDO05}
(where available).
Additionaly, for some objects, we took the black hole mass from \citet{GMM09b} and X-ray luminosity from \citet{GMM09}
and calculated Eddington ratios assuming typical LINER X-ray-to-bolometric correction $\kappa = 1.8$.
Then, the data were binned into sections of one decade in $L/L_{\rm Edd}$ and averaged.
Individual data points as well as the average spectral indices (and their dispersion) are shown in Fig.~\ref{fig12_corr}.
The average spectral index is consistent with a stable value $\sim 1.0$ and shows no correlation with luminosity. 

Our simulations show that the spectral slopes produced by a SSC mechanism are very stable.
Among all the parameters, the optical depth affects the slope most dramatically.
For a wide range of parameters changing by orders of magnitude ($\eta_{\rm B}=0.1-10$, $\tau = 0.1-3.0$, $\Gamma = 2-4$), the
corresponding spectral index variations are within $\Delta \alpha \lesssim 0.4$.
The typical values are $\sim$~0.8--1.0 and do not depend very much on the mass of the central object.
We also found that increasing the central mass at the same Eddington ratio leads to the asymptotic index $\alpha = 5/(s+2)$. 
For the steady-state index of the electron distribution $s = 3-5$, we have $\alpha \approx 0.7-1$, again consistent with the 
observed hard spectra.
This argues in favor of the SSC mechanism as the origin of the broad-band spectra in LINERs.

\subsection{Seyfert galaxies}
\label{sec:app_sy}

A typical X-ray spectrum of a Seyfert galaxy can be decomposed into an underlying power-law and a component arising from
reflection and reprocessing of the intrinsic emission. We are interested in formation of the intrinsic emission.
The low-luminosity end of Seyferts distribution at $L/L_{\rm Edd}\sim10^{-2}$ has intrinsic spectral slopes $\alpha\sim1$ very
similar to that of LINERs (see Fig.~\ref{fig12_corr}). 
Thus these sources can also well be described by pure SSC model.

We note, however, that Seyferts demonstrate spectral variations with varying flux. 
The dependence between the X-ray spectral slope and the X-ray flux for individual objects can often be fitted with a
power-law (\citealt{Chiang00}; \citealt*{DMZ00, Shih02}; \citealt{ChiBla03, Lamer03}). 
Similar correlations were found for GBHs, for example, in Cyg X-1 \citep{ZPP02}.
If we interpret the observed correlation in terms of the varying disc contribution $f(L)$ given by equation (\ref{eq:discfrac})
and assume that the spectral slope is determined by the accretion rate only (within small uncertainties corresponding to
specific parameters of the system), the data require $\beta \gtrsim 2$.
This implies a strong dependence of the inner radius of the accretion disc on the mass accretion rate.

More recently, \citet{Sob09} presented the statistical analysis of the correlation for ten AGNs. 
The average $\alpha-F$ correlation was shown to be less steep than that for each individual object.  
This effect might be caused by the fact that the same spectral slope in their data correspond to different Eddington ratios in
different objects. 
Thus tight correlation for any individual object is somewhat smeared out when considering an ensemble of sources. 
The reason of this smearing might be the badly determined mass of the central object. 
A similar slope--luminosity (in Eddington units) correlation is also observed in a 
large sample of Seyferts and quasars \citep{VF09,ZhouZhao10,ZZhang10} as shown in Fig.~\ref{fig12_corr}. 
We can interpret this correlation in terms of the varying contribution from the disc to the total luminosity 
and an increasing role of the disc photons for Comptonization. 
The dependence of the spectral slope on the Eddington ratio is shown in Fig.~\ref{fig12_corr}.
Two cases are considered: (1) varying optical depth
$\tau = \left(L/L_{\rm Edd}\right)^{1/2}$ with the disc luminosity fraction $f = 2 \left(L/L_{\rm Edd}\right)$;
and (2) the constant optical depth $\tau = 1.0$ with $f = 3 \left(L/L_{\rm Edd}\right)$ (the spectra and electron distributions
for this case are shown in Fig.~\ref{fig11_disc}).
In both cases strong correlation $\alpha - L/L_{\rm Edd}$ was found.
The best-fitting relation $\alpha \propto 0.3 \log \left( L/L_{\rm Edd}\right)$ found by
\citet{ZhouZhao10} matches well the corresponding lines from our simulations.
Thus at high luminosities $L/L_{\rm Edd}\gtrsim3\times 10^{-2}$ the observed spectral softening 
can be explained by the motion of the inner radius of the accretion disc towards the black hole and the 
increasing role of the accretion disc in supplying soft seed photons for Comptonization 
in the X-ray emitting region (corona or inner hot flow).

\subsection{Narrow line Seyfert 1 galaxies and quasars}
\label{sec:app_nlsy}

Narrow line Seyfert 1 galaxies seem to form a distinct class of Seyferts.
They are identified by narrow optical permitted and forbidden lines. 
The UV properties are unusual for presence of both low- and high-ionization lines.
Their X-ray spectra are on average steeper than those of Seyfert 1s \citep*{MDG07}, with commonly observed strong soft excess
and frequently detected high-amplitude variability.
The Eddington ratios are on average higher than in broad-line Seyfert 1s \citep[e.g.][]{Kom08}.
The power spectral density (PSD) spectrum of the NLSy1 NGC 4051 is similar to PSD of soft state of Cyg~X-1, but shifted to lower
frequencies \citep{McHPU04}.
Similar to the soft state of GBHs, one can expect that the two Seyfert classes differ by the presence of the accretion
disc close to the compact object.
If the disc extends very close to the SMBH, most of the energy is radiated in the UV range and completely
ionizes the medium around.
Hence, the lines in NLSy1s can be produced only in the region far away from the central object, where the gas velocities are
relatively low \citep{BFN94}. 

\begin{figure}
\centerline{ \epsfig{file=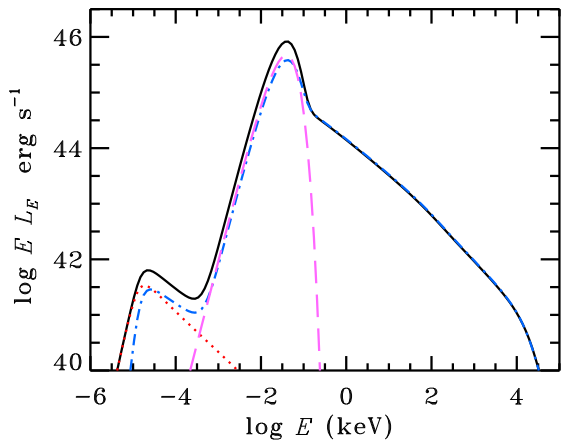,width=7.5cm}}
\caption{A hybrid Comptonization model for NLSy1s and quasars. 
The solid line corresponds to the total spectrum. Contribution from different processes: synchrotron (dotted line), Compton
scattering (dot-dashed line, sum of thermal and non-thermal), disc black-body (dashed line). 
The parameters are $M_{\rm BH} = 10^{8} \rm{M_{\odot}}$, $L/L_{\rm Edd} = 1.0$, $\tau = 1.0$, $\eta_{\rm B} = 0.25$, 
$\Gamma_{\rm inj} = 3.5$ and $f = 10$.
}
\label{fig13_excess}
\end{figure}

Soft excess is an ubiquitously observed feature in the 0.4--1 keV range in NLSy1s and radio-quiet quasars.
It can be modelled by cool Comptonization component \citep{PDO95, Vaug02}; however its position in the spectrum seems to be
independent on the central black hole mass, temperature of accreting disc and luminosity of the object. 
\citet{GD04_excess} proposed an alternative formation scenario where the excess arises from the smeared absorption of highly
ionized oxygen and iron, present in the hot wind from the accretion disc. 
For the quasar PG 1211+143, they obtained the slope of the intrinsic spectrum of $\alpha \approx 1.7$  in the range 0.4--10 keV.
This spectrum can be well reproduced by our model where the luminosity of the seed photons from the disc exceeds by ten times the 
power injected to electrons (see Fig.~\ref{fig13_excess}).
This results in a soft, power-law-like spectrum in the X-ray band produced by non-thermal Comptonization.
In this model, no cut-off is expected up to high ($\gg$ 100 keV) energies.

\section{Conclusions}

We have studied the spectral formation in hot non-thermal plasmas under the conditions relevant to the vicinities of SMBHs in
AGNs, using a self-consistent SSC model. 
We show that the SSC model can reproduce the exponential cut-off at energies above $\sim100$ keV 
and the power-law X-ray continuum with energy spectral indices $\alpha \sim 0.8-1.0$ 
observed in low-luminosity Seyferts and LINERs. 
The entire infrared/optical to X-ray spectrum in this objects can be produced by the SSC mechanism, suggesting a strong
correlation between the two energy bands.
No specific conditions are required to stabilize the slope for a wide range of parameters (magnetization, optical depth,
injection slope, Eddington ratio and black hole mass). 

We have demonstrated the difference in scaling of the SSC model and the two-phase disc-corona models. 
While in the latter case one would expect systematically softer spectra from objects with higher masses, in the case of SSC,
there exists a limiting slope (when the mass of the central object tends to infinity) $\alpha = 5/(s+2)$, which depends only on
the slope of the tail of the electron distribution. 
Comptonization spectra can get systematically softer  for objects of higher masses, if the electron injection function is 
sufficiently hard. 

We have tested the role of bremsstrahlung emission in SSC models.
We found that even though the process is not energetically significant, in certain cases, it leads to the hardening of the
power-law above a few keV and makes the overall X-ray spectrum concave.

We have also studied the role of additional soft blackbody photons from the accretion disc. 
With the increasing amount of soft photons, the equilibrium Maxwellian temperature drops and the X-ray spectrum softens.
The resulting spectra look similar to the GBHs in their soft state, which in the case of SMBHs are likely to be represented by
NLSy1s and quasars.
We have found that it is possible to explain the spectral slope -- flux correlation, widely discussed in the literature, 
by parametrizing the fraction of the disc photons in the medium as a power-law function of the Eddington ratio. 
This implies, that the inner radius of the truncated accretion disc is a strong function of the accretion rate.

\section*{Acknowledgments}

We thank the referee, Chris Done, for suggestions, which led to significant improvement of the paper.
This work was supported by the Finnish Graduate School in Astronomy and Space Physics (AV), the Wihuri
foundation and ERC Advanced Research Grant 227634 (IV), and the Academy of Finland grant 127512 (JP).


\label{lastpage} 

\end{document}